\documentclass[a4paper,11pt]{article}
\usepackage{setspace}
\usepackage[top=2.5cm,bottom=2.5cm, left=2.3cm, right=2.3cm]{geometry}
\usepackage{bm}
\usepackage{natbib}
\usepackage{graphicx}
\usepackage{epstopdf}
\usepackage{amsmath}
\usepackage{hyperref}
\usepackage{caption}
\usepackage{amsfonts}
\usepackage{theorem} 
\theoremstyle{break}

\title{Regression analysis with compositional data containing zero values}
\author{Michail Tsagris \\
Department of Computer Science \\
University of Crete, Heraklion, Greece \\ 
\href{mailto:mtsagris@yahoo.gr}{mtsagris@yahoo.gr}}

\begin{document}

\maketitle

\begin{abstract}
Regression analysis, for prediction purposes, with compositional data is the subject of this paper. We examine both cases when compositional data are either response or predictor variables. A parametric model is assumed but the interest lies in the accuracy of the predicted values. 
For this reason, a data based power transformation is employed in both cases and the results are compared with the standard log-ratio approach. There are some interesting results and one advantage of the methods proposed here is the handling of the zero values. 
\end{abstract}

\section{Introduction}
Compositional data are positive multivariate data whose vector elements sum to the same constant usually taken to be $1$ for convenience purposes. Data of this type arise in many disciplines, such as geology, ecology, archaeology, economics, geochemistry, biology, bioinformatics, political sciences and forensic sciences among others. In mathematical terms, their sample space, called simplex, is given by 
\begin{eqnarray} \label{simplex}
\mathbb{S}^d=\left\lbrace(x_1,...,x_D)^T \bigg\vert x_i \geq 0,\sum_{i=1}^Dx_i=1\right\rbrace, 
\end{eqnarray}
where $D$ denotes the number of variables (better known as components) and $d=D-1$. 

There are various techniques for regression analysis with compositional data being the response variables. See for example \cite{ait2003} who used classical methods on a log-ratio transformed space and \citet{stephens1982} and \citet{scealy2011} who transformed the data on the surface of a unit hyper-sphere using the square root transformation. Dirichlet regression models have been employed by \citet{gueorguieva2008} and \citet{maier2011}. 

When zero values exist in data, Dirichlet models and the log-ratio transformation suggested by \cite{ait1982, ait2003} and \citet{ilr2003} will not work unless a zero value imputation is applied first. The square root transformation on the other hand treats them naturally \citep{scealy2011}, but the procedure is not easy to implement. 

We propose the use of a newly suggested data based power transformation \cite{tsagris2011} to perform regression analysis. The multivariate logit link function is necessary to ensure that the fitted values lie within the simplex. The free parameter of the power transformation is chosen such that that the discrepancy between the observed and the fitted values is minimized. This implies that the use of Kullback-Leibler divergence will be of great importance in achieving this goal. 

The big advantage of this type of regression is that a) zeros are handled naturally and thus no zero imputation technique prior to the analysis is necessary and b) more accurate fitted values can be obtained. The disadvantage on the other hand is that we loose in terms of statistical properties of the estimated coefficients. Hence, this is more a data mining procedure. 

The inverse situation, where compositional data are in the predictor variables side has not been looked at, thoroughly, in the literature as far as we are aware of. One possible solution would be to apply the isometric log-ratio transformation \citep{ilr2003} (see \ref{ilr}) and then standard techniques, see for example \citet{egozcue2011} and \citet{hron2012}. But this approach has two drawbacks: first it does not account for any collinearity between the components and second it is not applicable when zero values are present. 

Collinearity problems can be attacked by principal component regression. Zero values require imputation, so that the isometric log-ratio transformation can be applied. This however means that the values of the rest components of each vector, which contains at least one zero value, will have to change. In some data sets, there can be many zeros spread in many observations. That would mean that many observations would have to change values, even slightly. This adds extra variation to the data though. 

In this paper we propose a novel parametric regression method whose ultimate purpose is prediction inference. A data based power transformation \citep{tsagris2011} involving one free parameter, $\alpha$ is employed at first. We suggest a way to choose the value of $\alpha$ which leads to the optimal results. The results show that prediction can be more accurate when one uses a transformation other than the isometric \citep{ilr2003} or the additive log-ratio \citep{ait1982}. 

A similar approach is exhibited when for the case of compositional data being predictor variables. Thus, this extends the work by \citet{hron2012} who used the isometric log-ratio transformation only. However, our work focuses mainly on prediction and not on inference regarding the regression coefficients. 

Alternatively, if the number of explanatory variables is rather small, standard linear regression analysis can be carried out. In either case, the $\alpha$-transformation handles zero values in the data naturally (if present) and principal component or $k$-NN regression handles possible collinearity problems. The key message is that a transformation other than the log-ratio family can lead to better results in terms of prediction accuracy. 

Regression analysis when compositional data are the response variables is described in Section 2. Simulation studies and a real data example illustrates the methodology. Section 3 describes the principal component regression when compositional data are the predictor variables. A real data example shows interesting results. Finally, conclusions close this paper.

\section{The $\alpha$-transformation for compositional data}
The $\alpha$-transformation was proposed by \cite{tsagris2011} as a more general than the isometric log-ratio transformation \citep{ilr2003} and is a data based power transformation involving one free power parameter, similarly to the Box-Cox transformation. The power transformation suggested by \citet{ait2003} is
\begin{eqnarray} \label{alpha}
{\bf u}=\left(\frac{x_1^{\alpha}}{\sum_{j=1}^Dx_j^{\alpha}}, \ldots, \frac{x_D^{\alpha}}{\sum_{j=1}^Dx_j^{\alpha}}\right),
\end{eqnarray}
\\
and in terms of that define
\begin{eqnarray} \label{isoalpha}
{\bf z}=\frac{1}{\alpha}{\bf H}\left(D{\bf u}-{\bf j}_D\right), 
\end{eqnarray} 
where ${\bf x} \in  \mathbb{S}^d$, ${\bf H}$ is the $d \times D$ Helmert sub-matrix (the Helmert matrix \citealp{helm1965} without the first row) and ${\bf j}_D$ is the $D$-dimensional vector of $1$s. Note that the power transformed vector ${\bf u}$ in (\ref{alpha}) remains in the simplex $S^d$, whereas ${\bf z}$ is mapped onto a subset of $R^d$. Note also that (\ref{isoalpha}) is simply a linear transformation of (\ref{alpha}) and moreover as $\alpha \rightarrow 0$, (\ref{isoalpha}) converges to the isometric log-ratio transformation \citep{ilr2003} defined as
\begin{eqnarray} \label{ilr}
{\bf v}={\bf H}\left(\log{x_1}-\frac{1}{D}\sum_{j=1}^D\log{x_j},\ldots,\log{x_D}-\frac{1}{D}\sum_{j=1}^D\log{x_j}\right)^T.
\end{eqnarray}

We can clearly see that when there are zero values in the compositional data the isometric log-ratio transformation (\ref{ilr}) is not applicable because the logarithm of zero is undefined. For this reason, we will also examine the zero value imputation briefly mentioned in the next Section.

\subsection{The $\alpha$-regression} \label{expalpha}
We will use the inverse of the additive logistic transformation, combined with the $\alpha$-transformation, as a link function. This is a new regression using the $\alpha$-transformation (\ref{isoalpha}) which allows for more flexibility even in the presence of zero values. Another feature of this method is that the line is always curved (unless $\alpha$ is far away from zero) and so it can be seen not only as a generalization of the log-ratio regression but also as a flexible type compositional regression in the sense that the curvature of the line is chosen based on some discrepancy criteria, examined later. 

In order for the fitted values to satisfy the constraint imposed by the simplex we model the inverse of the additive logistic transformation of the mean response. Hence, the fitted values will always lie within $S^d$ and we also retain the flexibility the $\alpha$-transformation offers. 

We assume that the conditional mean of the observed composition can be written as a non-linear function of some covariates
\begin{eqnarray} 
\begin{array}{cc} \label{regalpha}
\mu_1=\frac{1}{1+\sum_{j=1}^de^{{\bf x}^T\pmb{\beta}_j}} & \\
\mu_i= \frac{e^{{\bf x}^T\pmb{\beta}_i}}{1+\sum_{j=1}^de^{{\bf x}^T\pmb{\beta}_j}} & \text{for} \ \ i=2,...,D,
\end{array}
\end{eqnarray}
where 
\begin{eqnarray*} 
\pmb{\beta}_i=\left(\beta_{0i},\beta_{1i},...,\beta_{pi} \right)^T, \ i=1,...,d \ \ \text{and $p$ denotes the number of covariates}.
\end{eqnarray*}
 
Then a multivariate linear regression is applied to the $\alpha$-transformed data 
\begin{eqnarray} \label{loglikreg}
l\left(\alpha\right) &=& -\frac{n}{2}\log{\left|\hat{\pmb{\Sigma}}\right|}
-\frac{1}{2}tr\left[\left({\bf Y}_{\alpha}-{\bf M}_{\alpha}\right)
\hat{\pmb{\Sigma}}_{\alpha}^{-1}\left({\bf Y}_{\alpha}-{\bf M}_{\alpha}\right)^T \right],
\end{eqnarray}
where ${\bf Y}_{\alpha}$  and ${\bf M}_{\alpha}$ are the $\alpha$-transformed response and fitted compositional vectors. We have ignored the Jacobian determinant of the $\alpha$-transformation since it plays no role in the optimization process and the choice of $\alpha$ For each value of $\alpha$ we maximize the value of this objective function (\ref{loglikreg}). The $\hat{\pmb{\Sigma}}$ needs not be numerically estimated, since $\hat{{\bf B}}$, the matrix of the estimates and $\hat{\pmb{\Sigma}}$ are statistically independent \citep{mardia1979}. The maximum likelihood estimator of $\pmb{\Sigma}$ is \citep{mardia1979}
\begin{eqnarray*}
\hat{\pmb{\Sigma}}_{\alpha}=n^{-1}{\bf Y}_{\alpha}^T{\bf PY}_{\alpha}, 
\end{eqnarray*}
where ${\bf P}={\bf I}_n-{\bf X}\left({\bf X}^T{\bf X}\right)^{-1}{\bf X}^T$. But since this covariance is not unbiased we will use the unbiased estimator
\begin{eqnarray*}
\hat{\pmb{\Sigma}}_{\alpha}=\left(n-p-1\right)^{-1}{\bf Y}_{\alpha}^T{\bf PY}_{\alpha}, 
\end{eqnarray*}
where ${\bf X}$ is the design matrix and $p$ is the number of independent variables. 

The consistency of the estimators of the parameters is not an issue in our case since we focus on prediction inference. Since the estimation of the parameters depends upon the value of $\alpha$, the estimates will not be consistent, unless that is the true assumed model. The multivariate normal is defined in the whole of $\mathbb{R}^d$ but the $\alpha$-transformation maps the data onto a subset of $\mathbb{R}^d$. Thus, unless there is not too much probability left outside the simplex, the multivariate normal distribution might not be the best option. 

The $\alpha$-transformation what it does essentially is to contract the simplex, center it to the origin and then project it on a subspace of $\mathbb{R}^d$ by using the Helmert sub-matrix \citep{helm1965}. So if the fitted multivariate normal has high dispersion that will lead to probability left outside the simplex. The multivariate t distribution was used by \citet{lange1989} as a more robust, in comparison to the multivariate normal, model but even so, it will not be the best option, mainly for two reasons. Even if the multivariate t distribution could provide flatter tails, there would still be some probability (even less than the normal) left outside the simplex. Secondly, in a regression setting, the number of parameters we would have to estimate numerically is increased and this would make the maximization process more difficult. 

A final key feature we have to note is that when $\alpha \rightarrow 0$ we end up with the additive log-ratio regression (\ref{regalr}). 

\subsubsection{Choosing the optimal $\alpha$ using the Kullback-Leibler divergence}
The disadvantage of the profile log-likelihood, for choosing the value of $\alpha$, is that it does not allow zeros. On the other hand, it provides the maximum likelihood estimates which are asymptotically normal. Furthermore, confidence intervals for the true value of $\alpha$ can be constructed. 

We suggest an alternative and perhaps better way of choosing the value of $\alpha$. Better in the sense that it is trying to take into account the proximity between the observed and the fitted values. The criterion is to choose the $\alpha$ which minimizes twice the Kullback-Leibler divergence \citep{kullback1997}
\begin{eqnarray} \label{Mdiverge}
KL=2\sum_{j=1}^n\sum_{i=1}^Dy_{ij}\log{\frac{y_{ij}}{\hat{y}_{ij}}},
\end{eqnarray}
where $y_{ij}$ is the observed compositional point and $\hat{y}_{ij}$ is the corresponding fitted value. The form of the deviance for the log-linear models and the logistic regression has the same expression as well. Hence, we transfer the same form of divergence to compositional data. For every value of $\alpha$ we estimate the parameters of the regression and choose the value of $\alpha$ which minimizes (\ref{Mdiverge}). 

The number $2$ is there because in the case of $D=2$ we end up with the log-likelihood of the binary logistic regression. The Kullback-Leibler divergence (\ref{Mdiverge}) takes into account the divergence or the distance of each of the observed values from the fitted values.  

Since we are interested in prediction analysis we should use cross-validation to choose the value of $\alpha$. The reason why we did not choose to go down this road is because of time. The estimation of the parameters for a single value of $\alpha$ requires some seconds in a fine desktop. The search over many possible values of $\alpha$ requires a few minutes. Thus, even a $1$-fold cross validation would require a few hours and as the number of independent variables, and/or the sample size increase the computational time will increase as well. So, for the shake of speed we avoided this way. 

\subsection{Additive log-ratio regression}
The additive log-ratio transformation is defined as 
\begin{eqnarray} \label{alr}
z_i=\log\left(\frac{y_i}{y_D}\right) \ \ \text{for} \ \ i=1,\ldots,d,
\end{eqnarray}
where $d=D-1$, ${\bf y}_D$ is the last component playing the role of the common divisor, but by relabelling any component can play this role. We will now show the additive log-ratio regression. At first we will see a nice property of the logarithms and its implications on the additive log-ratio regression.
\begin{eqnarray} \label{regalr}
\log\left(\frac{y_i}{y_D}\right)={\bf x}^T\pmb{\beta}_i \Leftrightarrow
\log{y_i}=\log{y_D}+{\bf x}^T\pmb{\beta}_i, \ \ i=1,\ldots, d
\end{eqnarray}
where ${\bf x}^T$ is a column vector of the design matrix ${\bf X}$, $D$ is the number of components and
\begin{eqnarray*}
\pmb{\beta}_i=\left(\beta_{0i},\beta_{1i},...,\beta_{pi} \right)^T, \ i=1,...,d
\end{eqnarray*}
are the regression coefficients and $p$ is the number of independent variables. 

We see from (\ref{regalr}) that when the dependent variable is the logarithm of any component, the logarithm of the common divisor component  can be treated as an offset variable; an independent variable with coefficient equal to $1$.   

The main disadvantage of this type of regression is that it does not allow zero values in any of the components, unless a zero value imputation technique \cite{martin2012} is applied first. Its advantage though is that after the additive log-ratio transformation (\ref{alr}) we can do the standard multivariate regression analysis. As for the fitted value they are back transformed into the simplex using the inverse of (\ref{alr}).

\subsection{Example: Foraminiferal compositions with some zero values} \label{foraminifers}
This data set consists of foraminiferal (marine plankton species) compositions at $30$ different depths ($1$-$30$ metres) and can be found in \cite[pg.~399]{ait2003}. There are $5$ compositions with a zero value, either in the third or the fourth component. The data were analysed by \cite{scealy2011} who performed regression by employing the Kent distribution using the logarithm of the depth as the independent variable. 
Since the data contain zero values, the $\alpha$-regression allows only strictly positive values for $\alpha$. Since the composition belongs to $S^3$ we cannot use the ternary plot; we could though use a $3$-dimensional pyramid, but it would not show the structure of the data clearly. For this reason we will use a barplot.

\begin{figure}[ht]
\centering
\includegraphics[scale=0.5,trim=50 20 50 50]{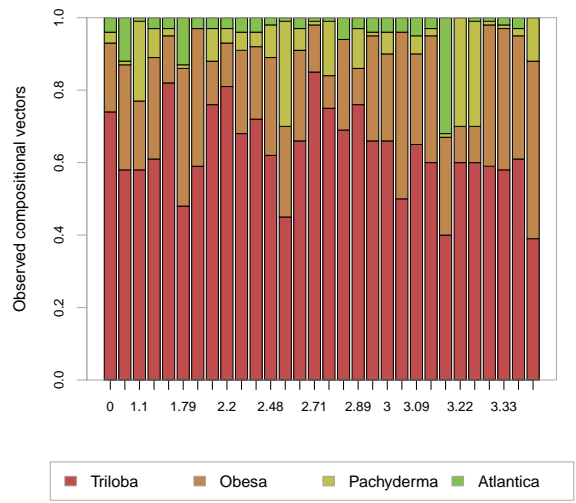}
\caption{The foraminiferal compositions as a function of the logarithm of the sea water depth.}
\label{reg1}
\end{figure}

A zero value imputation method suggested by \citet{martin2012} can be briefly summarised as follows: Replace the zero values by $65\%$ of a threshold value using the non parametric method described by \citep{martin2003}. The threshold value is different for each component. We used the minimum, non zero, value of each component as that threshold. Then the isometric log-ratio transformation (\ref{ilr}) is applied. An E-M algorithm substitutes the replaced values with a value less than the chosen threshold. In the end, the data are transformed back to the simplex. This method, like all zero value imputation methods, results in all the components of each compositional vector being changed, even slightly. A tolerance value (say $0.01$) is used as convergence criterion between successive iterations. 

We will use this method in order to allow for the additive log-ratio regression to be performed. In this case the Kullback-Leibler divergence (\ref{Mdiverge}) will be calculated for the initial, not the zero imputed data.

\begin{figure}[!ht]
\centering
\begin{tabular}{cc}
\includegraphics[scale=0.5,trim=0 20 0 20]{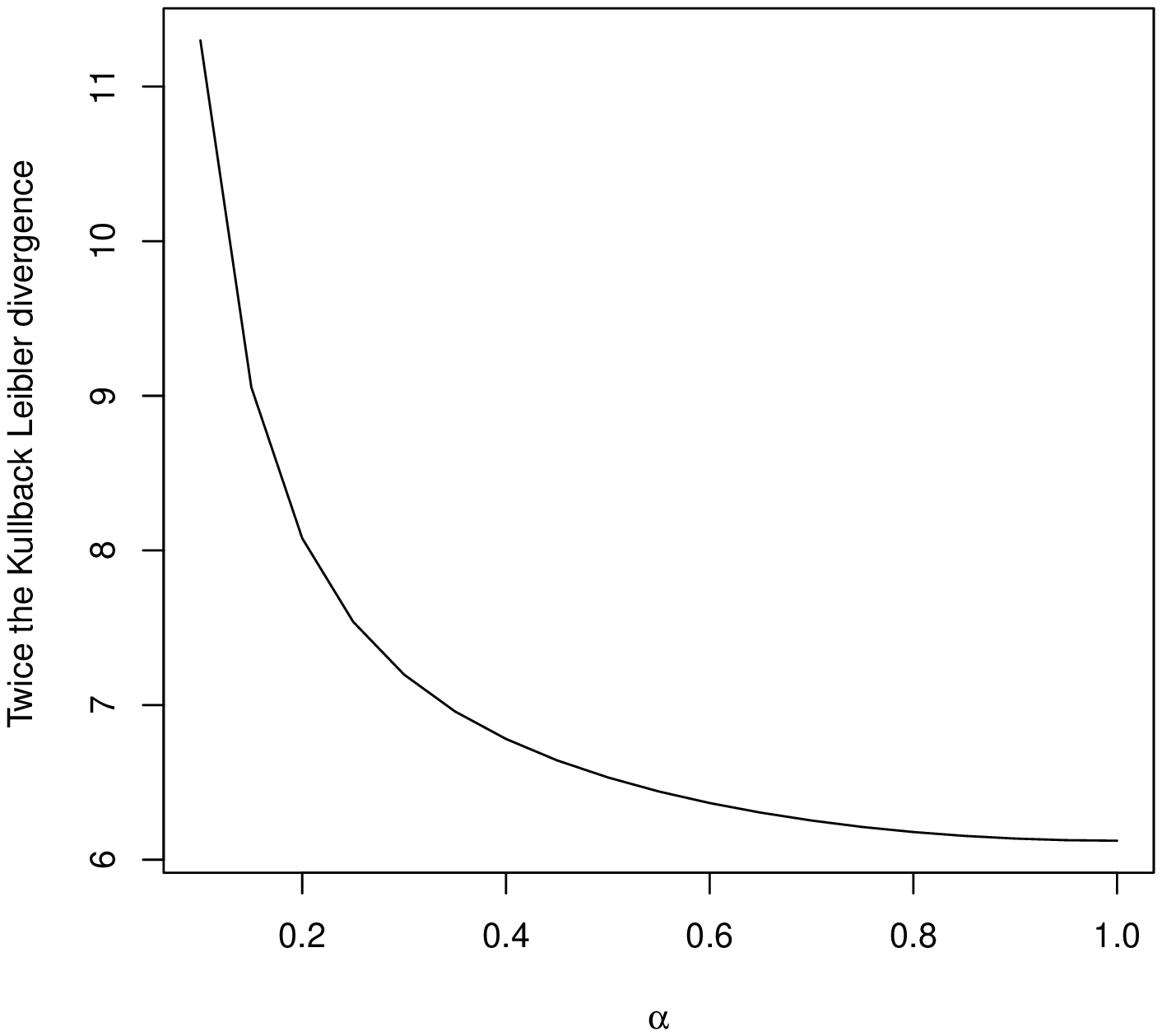} &
\includegraphics[scale=0.5,trim=0 20 0 20]{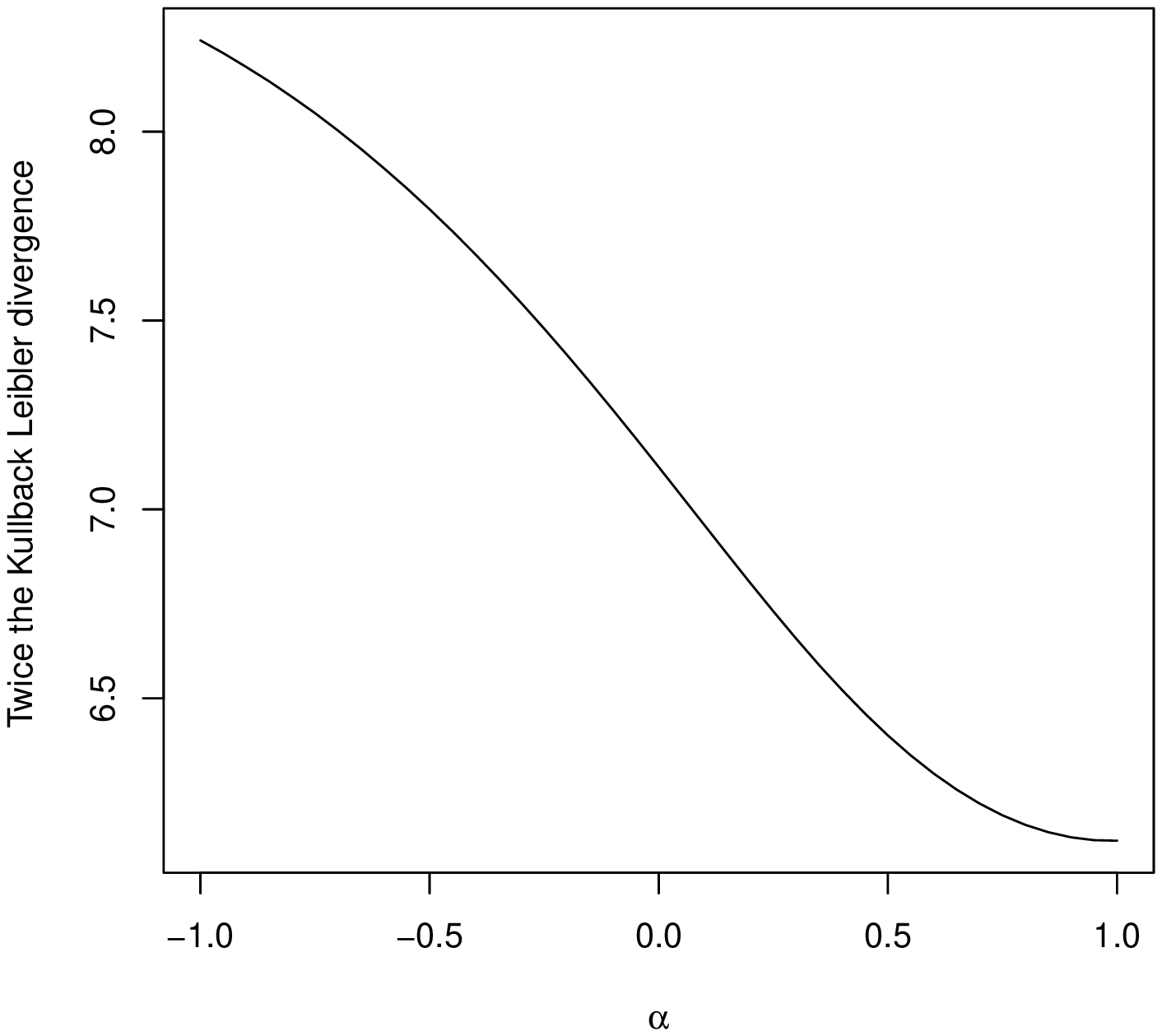}    \\
\footnotesize{(a)}   &  \footnotesize{(b)}   
\end{tabular}
\caption{Twice the Kullback-Leibler divergence (\ref{Mdiverge}) as a function of $\alpha$. Graph (a) refers to the original data and graph (b) refers to the zero imputed data.}
\label{kl}
\end{figure}

The value of the divergence (\ref{Mdiverge}) when regression is applied to the original data with $\alpha=1$ is equal to 6.123. The value of the divergence (\ref{Mdiverge}) when regression is applied to the zero imputed data data with $\alpha=0$ is equal to 7.112 and when $\alpha=1$ is equal to 6.123. Thus a value of $\alpha$ other than zero leads to better predictions with without zero value imputation. \citet{scealy2011} used the same dataset treating the compositional data as directional by applying the square root transformation. Their suggested Kent regression (no zero imputation is necessary) produced a divergence value equal to 6.344, still lower than the additive log-ratio regression. 

As earlier stated, the purpose of this regression is to provide better fittings to the observed data. It is more a data mining procedure than it is a statistical one, from the point of view that we are interested in estimating/predicting future observations with as high accuracy as possible. We are not interested in making inference about the regression coefficients. 

In the next Section, the reverse scenario is of interest, that of the compositional data (containing zero values again) playing the role of the covariates. 

\section{Principal component regression with compositional data being the predictor variables} \label{pcr}
As the title of this Section suggests we will focus only in the principal component regression, even though we could mention regression analysis in general. The  reason for not doing so is because \citet{hron2012} has already covered this case and we are interested in generalising this idea to cover the case of multicollinearity as well. 

Principal component regression is based on principal component analysis and hence we will not spend too much time. The algorithm to perform principal component regression (PCR) can be described as follows

\begin{enumerate}
\item At first standardize the independent variables. This way, ${\bf X}^T{\bf X}$ (the $n \times p$ design matrix, which includes the $p$ independent variables but not the intercept term) is proportional to the correlation matrix for the predictor variables \citep{jolliffe2005}. The $n$ stands for the sample size. 
\item Perform eigen analysis on ${\bf X}^T{\bf X}$ and calculate the matrix of the eigenvectors ${\bf V}$ and the scores ${\bf Z}={\bf XV}$. 
\item Estimate the regression coefficients by 
\begin{eqnarray*}
\hat{\bf B}={\bf V}\left({\bf Z}^T{\bf Z}\right)^{-1}{\bf Z}^T{\bf y},
\end{eqnarray*}
where ${\bf y}$ is the vector containing the values of the dependent variable. 
\item Estimate the covariance matrix of the estimated regression coefficients by
\begin{eqnarray*}
Var\left(\hat{\bf B}\right)=\sigma^2{\bf V}\left({\bf Z}^T{\bf Z}\right)^{-1}{\bf V}^T,
\end{eqnarray*}
where $\sigma^2$ is the conditional variance of the dependent variable calculated from the classical multiple regression analysis based upon the given number of principal components. It is the error variance, whose estimate is the (unbiased) mean squared error.  
\end{enumerate}

The key point is that we can have $p$ different sets of estimated regression coefficients, since we can use the first eigenvector (or principal component), the first two eigenvectors or all of them. If we use all $p$ of them, then we end up with the same regression coefficients as if we performed a classical multiple regression analysis. 

Since we are not interested in statistical inference regarding the coefficients of the principal components we can ignore their variance matrix. Another key point that is worthy to mention is that we can also use the principal scores as the new predictor variables and do the classical regression. The fitted values will be exactly the same. 

The idea here is simple, apply the $\alpha$-transformation (\ref{isoalpha}) to the compositional data and then perform principal component regression. The value of $\alpha$ and the number of principal components which lead to the optimal results are chosen via cross validation. The optimal results in our case will refer to minimization of the mean squared error of prediction. 

\subsection{Example: Glass data and refractive index (zero values present)} \label{glass}
In optics the refractive index of a substance (optical medium) is a "clean" number that describes how light, or any other radiation, propagates through that medium. It is defined as $\text{RI}=\frac{c}{v}$, where $c$ is the speed of light in vacuum and $v$ is the speed of light in the substance. The refractive index of water, for example, is $1.33$. This means that the speed of light in water is reduced by $24.8\%$ $\left[\left(1-\frac{1}{1.33}\right)\%\right]$. 

Surprisingly enough, negative refractive values are also possible, when if permittivity and permeability have simultaneous negative values. This can be achieved with periodically constructed metamaterials. The negative refraction index (a reversal of Snell's law) offers the possibility of the superlens and other exotic phenomena.

The glass dataset available from \href{http://archive.ics.uci.edu/ml/datasets/Glass+Identification}{UC Irvine Machine Learning Repository} contains $214$ observations regarding glass material from crime scenes. with information about $8$ chemical elements, in percentage form. These chemical elements are sodium (Na), Magnesium (Mg), aluminium (Al), silicon (Si), potassium (K), calcium (Ca), barium (Ba) and iron (Fe). The variable whose values we wish to predict based on knowledge of the chemical composition is the refractive index (RI) of each glass. 

The category of glass is also available (for example vehicle headlamps, vehicle window glass, tableware et cetera). This extra information can be taken into account by principal component regression easily. For the kernel regression though, an extra kernel for this discrete variable has to be fitted and the final kernel is the product of the two kernels (one for the continuous variables and one for this discrete variable). So, the inclusion of non continuous or different types of continuous data is not an easy task when kernel regression is to be used.  

We have also included a second PCR model which includes the extra information about the data, the category of the glass. The difference from the PCR described  is that now we will use the scores of the principal component analysis (see the second key point in Section \ref{pcr}) to do regression. In this way we will also add the categorical variable indicating the category of each glass measurement. Table \ref{tab1} summarizes the results of the five regression models. 

\begin{table}[h]
\begin{small}
\begin{center}
\begin{tabular}{cccc} 
                     &  \multicolumn{2}{c}{Regression model for the original data}    &        \\ \hline
Method               & Chosen parameters        & MSPE   & Adjusted $R^2$   \\ \hline  \hline
PCR                  & $\alpha=1$ and 7 PCs     & 1.237  & 0.891            \\  \hline
PCR with the glass   &                          &        &                  \\     
category             &  $\alpha=1$ and 7 PCs    & 1.02   & 0.903            \\  \hline 

                     &  \multicolumn{3}{c}{Regression model for the zero value imputed data}  \\ \hline
Method               & Chosen parameters        & MSPE   & Adjusted $R^2$   \\ \hline  \hline
PCR                  & $\alpha=0$ and 7 PCs     & 2.403  & 0.784            \\ \hline 
PCR                  & $\alpha=0.95$ and 7 PCs  & 1.239  & 0.890            \\ \hline
PCR with the glass   &                          &        &                  \\     
category             &  $\alpha=1$ and 6 PCs    & 1.016  & 0.863            \\  \hline         
\end{tabular}       
\caption{Mean squared error of prediction (MSPE) for each regression model applied to the glass data. The adjusted coefficient of determination ($R^2$) is calculated using the fitted values from principal component regression.}
\label{tab1}
\end{center}
\end{small} 
\end{table}

We can see from Table \ref{tab1} that even if we replace the zero values the result is the same. The isometric log-ratio transformation when applied to the zero imputed data does not lead to the optimal results. A value of $\alpha$ far from zero led to much better results, the adjusted $R^2$ increased from $0.772$ to $0.89$ and the mean squared error of prediction was less than half. \citet{martin2012} also used a robust E-M algorithm by employing MM estimators \citep{maronna2006}. The results (not presented here) when robust imputation was performed were slightly worse than the ones presented in Table (\ref{tab1}). 

Even if we impute the zero values, the estimated MSPE, when the isometric log-ratio transformation is used ($\alpha=0$), is much higher than with a value of $\alpha$ close to $1$ and the value of the adjusted $R^2$ is lower. 

\begin{figure}[!ht]
\centering
\begin{tabular}{ccc}
\includegraphics[scale=0.3,trim=0 20 0 20]{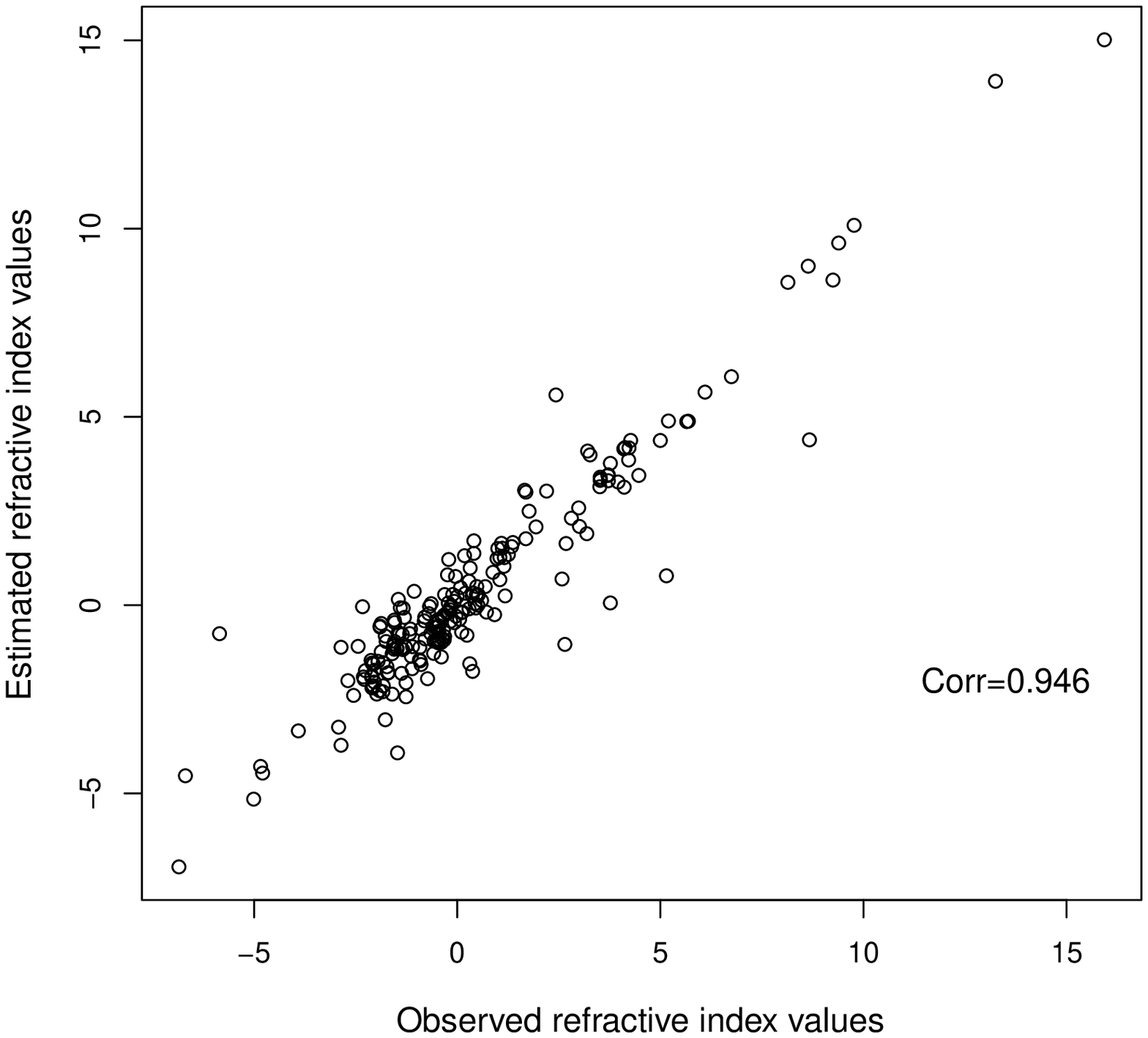} &
\includegraphics[scale=0.3,trim=0 20 0 20]{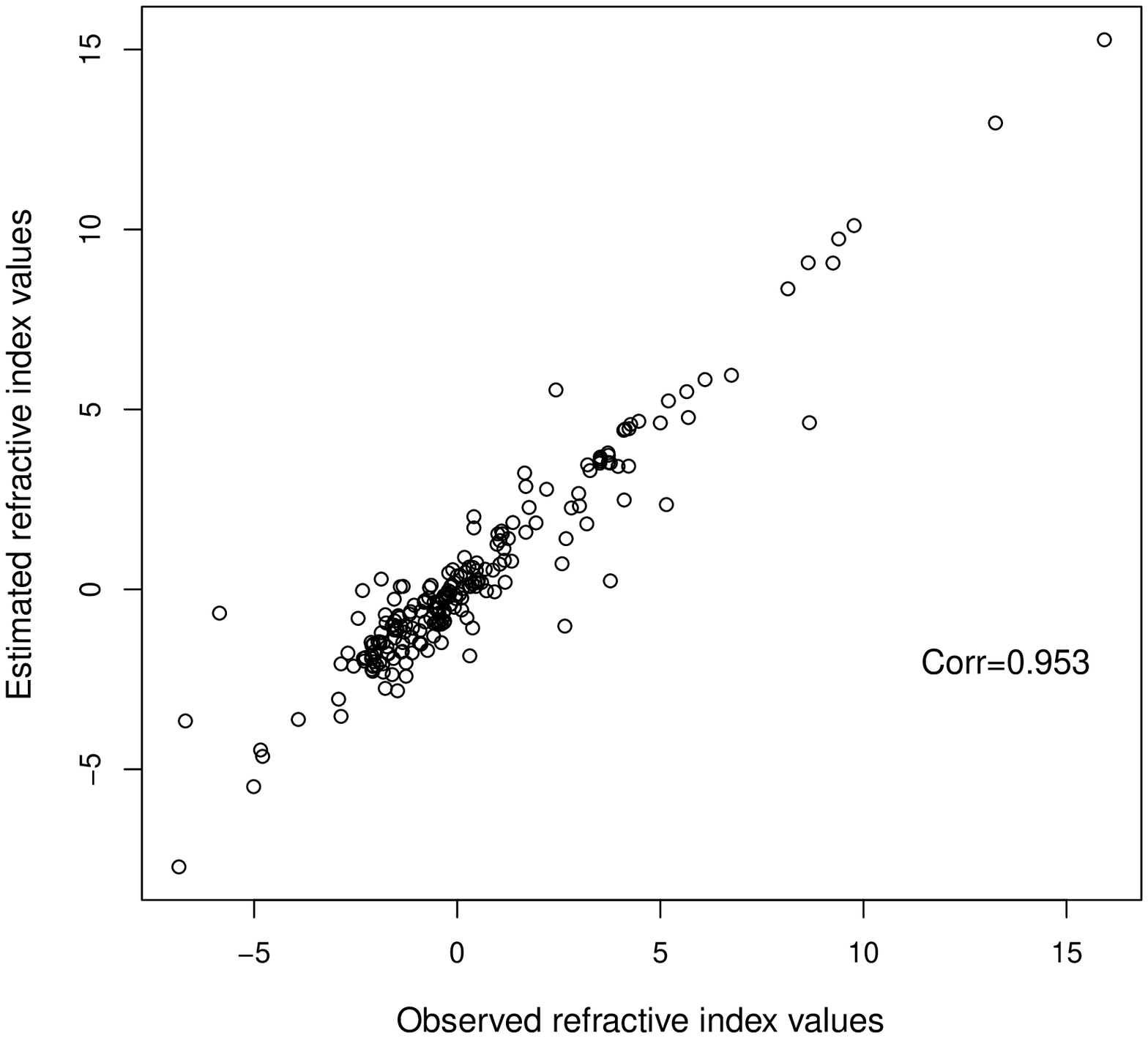} &
\includegraphics[scale=0.3,trim=0 20 0 20]{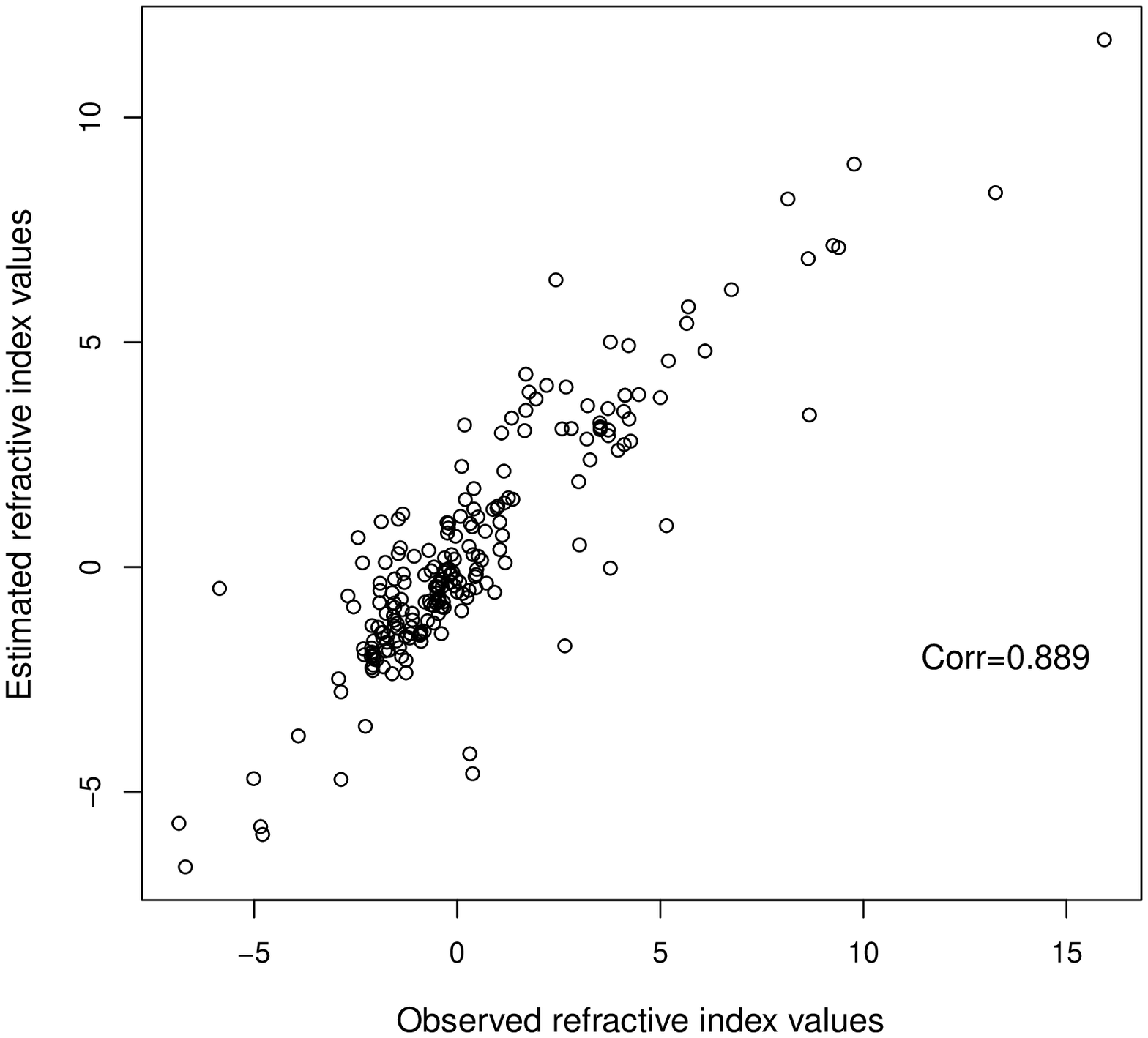}  \\
\footnotesize{(a)}   &  \footnotesize{(b)}   &  \footnotesize{(c)} \\
\end{tabular}
\caption{Observed versus estimated refractive indices. Plots (a) and (b) correspond to $\alpha=1$ with 7 PCs without and with information about the glass category, using in the original data. The third plot (c) corresponds to $\alpha=0$ with 7 PCs using the zero imputed data.}
\label{fig1}
\end{figure}

We also used kernel regression and $k$-NN regression using many different metrics or distances with and without the $\alpha$-transformation but none of them was as successful as the principal component regression. We also tried robust principal component regression for the second example but the results were not better either. 

The best model is the one using all 7 principal components coming from the $\alpha$-transformed data with $\alpha=1$ and using the information about the category of glass as we can see in Table \ref{tab1}. In mathematical terms it is written as
\begin{eqnarray*}
\hat{RI} &=& 1.345-0.081S_1+1.368S_2+0.161S_3-0.801S_4+2.533S_5-1.618S_6-1.329S_7 \\
         & & -1.567\text{WinF}-1.460\text{WinNF}-2.445\text{Veh}-1.165\text{Con}-1.177\text{Tabl}, 
\end{eqnarray*}
where $S_i$, for $i=1,\ldots,7$ stands for the scores of each principal component and WinF and WinNF stand for the window float and window non-float glass respectively. Veh stands for vehicle window glass, Con for containers, Tabl for tableware. The vehicle headlamps is the reference glass category. We do not show the standard errors since the choice of the number of principal components, and the inclusion of the information about the glass categories was based on the MSPE (Table \ref{tab1}). 

\begin{figure}[!ht]
\centering
\includegraphics[scale=0.4,trim=0 20 0 20]{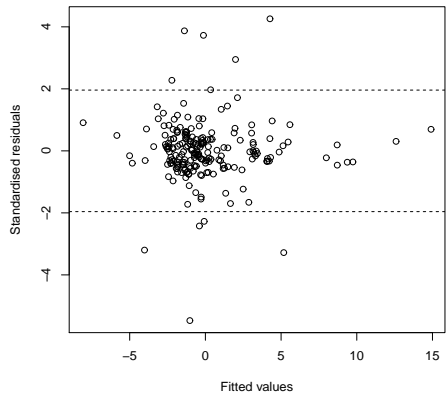} 
\caption{Fitted values versus the standardised residuals for the model. The two horizontal lines indicate the region of outliers in the residuals space.}
\label{fig2}
\end{figure}

The normality of the residuals was rejected according to the Shapiro test (p-value$<0.001$). As for the independence assumption by looking at Figure \ref{fig2} we can see that it looks acceptable. There are a few outliers in the residuals space as we can see from Figure \ref{fig2}. The $\alpha$ used in this model was equal to $1$ as we mentioned before. Different values of $\alpha$ will lead to different values bering detected as potential outliers. In addition, we detected some outliers in the residuals space and not in the space spanned by the $7$ principal components. 

\section{Conclusions}
There are two key messages this paper tries to convey. The first is that a transformation other than the isometric of the additive log-ratio transformation should be considered for compositional data analysis. It was evident from the example that when the data contain zero values the $\alpha$-transformation handles the data naturally without changing their values even to the slightest. The log-ratio transformations require zero value imputation prior to their applications,  a requirement not met by the $\alpha$-transformation. 

One can argue though that in both examples values of $\alpha$ close to 1 produced similar results as $\alpha=1$. This was something to be expected for the data in the first example, but even then, this discourages the use of log-ratio as the panacea for all compositional data analysis. We do not disqualify though the use of log-ratio but rather try to show that this very popular transformation for compositional data has some drawbacks and in some cases other transformations could be applied. In both examples, the differences in the Kulback-Leibler divergence or the MSPE were small, but still indicative. 

The second key message is that when compositional data analysis are on the covariates side standard regression techniques after the $\alpha$-transformation should be used with caution. Since the data represent percentage allocation, it is very natural that correlations exist even after the transformation and thus, multicollinearity should be examined first. In addition, if there are many components then variable selection could be applied. These are some reasons why we preferred to perform principal component regression. 

\section*{Acknowledgements}
The author would like to express his acknowledgments to the anonymous reviewer for his comments.

\end{document}